\documentclass{aastex}
\usepackage{emulateapj5,epsfig}
\usepackage{epsfig}

\begin{document}
\title{
Effects of Early Cosmic Reionization \\
on the Substructure Problem in Galactic Halo
}
\author{Hajime Susa\altaffilmark{1} 
\vskip 0.2cm
\affil{Department of Physics, Rikkyo University, Nishi-Ikebukuro,
Toshimaku, Japan}
\vskip 0.3cm
Masayuki Umemura\altaffilmark{2}
\vskip 0.2cm
\affil{Center for Computational Sciences, University of
  Tsukuba, Japan }
\altaffiltext{1}{susa@rikkyo.ac.jp}
\altaffiltext{2}{umemura@rccp.tsukuba.ac.jp}
}

\begin{abstract}
Recent observations on the cosmic microwave background 
by {\it Wilkinson Microwave Anisotropy Probe} ({\it WMAP}) 
strongly suggest that 
the reionization of
the universe took place quite early ($z\sim 17$). On the other hand, 
it has been pointed out that the cold dark matter
cosmology suffers from the substructure problem 
that subgalactic halos are overproduced than the observed dwarf
galaxies in the Local Group. In this paper, as a potential mechanism 
to solve this problem, 
we consider the feedback effects of the early reionization on
 the formation of small-scale structures.
For this purpose, we perform 3D radiation hydrodynamic simulations
with incorporating the radiative transfer for ionizing photons. As a result, 
it is found that the early reionization is so devastating for low-mass systems
with $M_{\rm vir} \la 10^8 M_\odot$ or $v_{\rm circ}\la 20{\rm km ~ s^{-1}}$,
and almost all gas is photo-evaporated in more than 95\% of low-mass systems. 
Such a strong negative feedback on
 the formation of low-mass galaxies may solve the substructure problem
and support the picture that
 Local Group dwarf galaxies are descendants of the more massive halos
 that experienced and survived tidal stripping .
\end{abstract}


\section{Introduction}
\label{intro}
It has been claimed that too many subgalactic halos are formed in a
cold dark matter (CDM) universe, compared to the dwarf galaxies observed in the Local Group
 \citep{Klypin99,Moore99}. 
One of the simplest ways to resolve this discrepancy is the exclusion
of baryonic matter from small halos by some feedback mechanisms.
The mechanisms could be either the gas ejection due to energy
input by the multiple supernova (SN) explosions
\citep{DS86,Yepes97,Efstathiou00,Kay02,Marri02,Wada03,Ricotti04}
 or 
the photo-evaporation by ultraviolet background (UVB) radiation
\citep{UI84,BSS88,Efstathiou92,BR92,TW96,BL99,KI00,Kitayama00,Kitayama01,Bullock00,Somerville02,Benson02,Ricotti02,SU04}. 
As for the former process, \citet{Wada03} recently 
performed high resolution hydrodynamic simulations 
and concluded that 1000 SNe are required in order to disrupt a galaxy 
with $M\sim 10^8 M_\odot$ at $z\sim 10$, while only 100 SNe in
the halo lead to triggering the re-collapse of 
the whole system. These results show that SN explosions are not always
destructive for the formation of low-mass galaxies, although it depends
upon the initial mass function of stars (e.g. Ricotti \& Ostriker 2004).
On the other hand, the previous studies on the UVB feedback 
have shown that the photo-evaporation by UVB is dependent upon 
the virial mass. 
As long as systems are more massive than
$\approx 10^8M_\odot$, 
the photo-evaporation by UVB is not effective,
because deep potential wells can retain the ionized gas.
However, at smaller mass-scales, the feedback by UVB
is expected to play a very important role. 
Recently, {\it Wilkinson Microwave Anisotropy Probe} ({\it WMAP}) \citep{Kog03}
suggested that
the universe was reionized in a rather early epoch ($z \sim 17$). 
If this is the case, the UVB feedback could be quite effective for low-mass
systems, since density fluctuations are photo-heated 
before they collapse to form stars.

Very recently, \citet{Kravtsov04} analyzed the dynamical history 
of small substructure halos with present mass $\la 10^{8-9} M_\odot$
by a cosmological simulation. They found that 10\% of small halos
originate in considerably larger systems with $\ga 10^9 M_\odot$,
which survived the tidal stripping. 
They suggest that the Galactic satellites are descendants of relatively 
massive systems which formed at higher redshifts and 
did not significantly suffer from the UVB.
This argument is plausible, if systems with $\ga 10^9 M_\odot$ 
are impervious even to the early reionization, and also
the star formation in low-mass systems is completely suppressed.

\cite{Kitayama01} have investigated the UVB
feedback by one dimensional
radiation hydrodynamic simulation for halos which collapse at 
$z\la 10$, 
and found that the central parts cannot cool down to form stars if
the virial temperature is less than $10^4$ K and $I_{21}\ga 10^{-2}$,
where $I_{21}$ is the intensity at Lyman limit in units of 
$10^{-21}$ erg s$^{-1}$ cm$^{-1}$ Hz$^{-1}$ str$^{-1}$.
\citet{Dijkstra04} extended this work to higher redshifts
by one dimensional hydrodynamic calculations, and 
found that gas is not photo-evaporated in halos 
with circular velocities of $10-20{\rm km~s^{-1}}$ at $z>10$. 
But, in the hierarchical structure formation in a cold dark matter 
universe, three-dimensional (3D) time-dependent self-shielding 
is significant for the UVB feedback, as shown by \citet{SU04}.
In particular, such radiative transfer effects are likely to be
essential, when the reionization proceeds at
higher redshifts (in higher density intergalactic matter).

In this paper, we perform 3D radiation
hydrodynamic simulations with solving the radiative transfer 
for ionizing photons, and attempt to elucidate the feedback 
effects of the early reionization on the formation of 
subgalactic systems. 
In the next section, the method of numerical simulations
is briefly described, and the assumption for ionizing radiation intensity
is provided. In \S \ref{Results}, we present the numerical results for 
UVB feedback and also make the convergence check of the results.
\S \ref{Discussion} is devoted to the discussion on the substructure problem.

\section{Numerical Method}
\label{numsim}
The details of numerical method is given in \citet{SU04}. 
Here, we briefly describe the model and method for simulations.

Hydrodynamics is calculated by Smoothed Particle Hydrodynamics (SPH)
method. We use the version of SPH by \citet{Ume93} with the modification
according to \citet{SM93}, and also we adopt the particle resizing
formalism by \citet{Thac00}. 
The gravitational force is calculated by a special purpose processor for
gravity, GRAPE-6 \citep{Makino02}. 
In order to access GRAPE boards, we utilize the Heterogeneous
Multi-Computer System (HMCS) \citep{HMCS} which allows us to use 
GRAPE in parallel processors such as PC clusters. 

The softening length for gravity is set to be $20$pc 
for SPH and CDM particles. This value is based on the
convergence test shown in section \ref{Results}.
The non-equilibrium chemistry and radiative cooling 
for primordial gas are included with the code
developed by \citet{SuKi00}, where H$_2$ cooling and
reaction rates are taken from \citet{GP98}.
To solve radiative transfer in an SPH scheme, we employ
the method proposed by \citet{KB00}, which utilizes the neighbour lists
of SPH particles to evaluate the optical depth from a certain source to 
an SPH particle. 

To generate UV background radiation, we put a single 
ionizing source located very far 
from the simulated region, and control the UV intensity 
by specifying the incident flux to the simulation box.  
We assume the history of UV intensity that allows
the early reionization inferred by {\it WMAP}.
Since the recombination time scale is
shorter than the Hubble expansion time at $z\ga 10$
\citep{Peebles68}, ionizing
photons should be continuously supplied in order to retain the ionization.
The requisite UV intensity is given by equating the recombination rate 
to ionization rate:
\begin{equation}
I_{\nu_{\rm L}}=\frac{n_{\rm 0}(1+z)^3
k_{\rm rec}h_{\rm P}(\alpha+3)}{4\pi\sigma_{\nu_{\rm L}}}
\frac{(1-y_{\rm HI})^2}{y_{\rm HI}},
\label{eq:I21}
\end{equation}
where $I_{\nu_{\rm L}}$ denotes the intensity at Lyman limit, and
$k_{\rm rec}$, $\sigma_{\nu_{\rm L}}$, and $h_{\rm P}$ respectively 
represent the recombination coefficient, ionization cross-section 
at Lyman limit, and Plank constant.  
$n_0$ is the present-day baryon density,  
$\alpha$ is the spectral index of UV radiation,  
and $y_{\rm HI}$ is the fraction of neutral hydrogen.
If $y_{\rm HI}=0.1$ at $z\simeq 15$, $I_{21}\simeq 10^{-3}$ is required.
This should be regarded as the minimal prerequisite for two
reasons.
First, this estimate is provided for optically-thin media.
In actual ionization history, the radiative transfer effect is 
definitely important \citep{NUS01}.
Secondly, in an inhomogeneous universe, local density enhancements 
increase the recombination rate significantly. 
Taking these into account, we assume $I_{21}=0.01$ for $5<z<17 $
with a sharp decrease at $z>17$ as $I_{\rm 21}\propto
\exp\left[3\left(17-z\right)\right]$.
Also, some models with $I_{21}=1$ are simulated to 
investigate the dependence on $I_{21}$.

The ``star formation'' condition adopted in this paper is basically
the same as \citet{SU04}, except that $c_*=1$ is assumed here. 
$c_*=1$ means the star formation at maximal feasible rate, 
because no stars can form in a timescale shorter than local free-fall time.  
Here, we do not include the internal feedback effects by
stellar UV radiation as well as SN explosions.

We assume $\Omega_{\rm M}=0.3$, 
$\Omega_{\rm \Lambda}=0.7$, 
$\Omega_{\rm baryon}h^2=0.02$, and $h=0.7$, as the background cosmology. 
Following the density fluctuations in this cosmology, 
the initial distributions of particles are set up.
The mass of virialized dark halo is in
the range of $10^6 M_\odot \la M_{\rm vir} \la 10^8 M_\odot$, 
and the collapse epoch is $5 \la z_{\rm c} \la 20$.
We use $2^{15}$ SPH particles and $2^{15}$ dark matter particles 
for a run.

\section{Results}
\label{Results}
The numerical results are summarized in Figure \ref{fig1}.
The left panel shows the fraction of the final mass in stellar component
to the initial baryon mass as a function of the collapse epoch 
$z_{\rm c}$ and $M_{\rm vir}$, while the right panel is
the final fraction of gas. 
In both panels, the fraction is depicted by different symbols.
Dotted lines
represent the collapse epoch of halos formed from $1\sigma, 2\sigma$ and
$3\sigma$ CDM density fluctuations. Roughly 95\% of fluctuations collapse
after the epoch predicted by the $2\sigma$ line. 
Dashed lines denote the constant circular velocities,
which are defined by $v_{\rm circ}\equiv \sqrt{GM_{\rm vir}/r_{\rm vir}}$
with the virial radius $r_{\rm vir }$ determined by $z_{c}$ and $M_{\rm vir}$.

In Figure \ref{fig1}, we see that, 
if $M_{\rm vir} \ga 10^8M_\odot$ and $v_{\rm circ}\ga 20{\rm km ~ s^{-1}}$,
almost all baryons are transformed into stars, leaving little gas. 
It is noted that a considerable fraction of baryonic matter form stars even
after the reionization  ($z_{\rm c}<z_{\rm reion}=17$).
On the other hand, if $v_{\rm circ}\la 20{\rm km ~ s^{-1}}$,
$f_{\rm star}$ steeply decreases with decreasing circular velocities.
Also, $f_{\rm gas}$ becomes quite small as seen in the right panel.
This reflects the effect of photo-evaporation by the UVB.
The dependence of photo-evaporation on $M_{\rm vir}$ and $z_{\rm c}$ 
is understood by the self-shielding and the gravitational potential. 
The self-shielding against UVB becomes prominent
if the local density exceeds a threshold density as 
\begin{equation}
n_{\rm shield}=1.4\times10^{-2} {\rm cm}^{-3} 
\left(M_{\rm baryon} \over 10^8M_{\rm {\odot}}\right)^{-1/5}
\left( I_{21} \over \alpha \right)^{3/5} \label{nshield}
\end{equation} 
\citep{TU98}. This means that the self-shielding is more
effective if the system is larger or collapses at higher redshifts.
The gas envelope that is not shielded from UVB is
photo-heated to $\la 10^4$K and is blown out by 
the enhanced thermal pressure, unless the
gravitational potential is deep enough to retain the ionized gas
(i.e. $v_{\rm circ}\ga 20{\rm km ~ s^{-1}}$). 
Hence, the photo-evaporation is quite devastating 
for fluctuations with lower masses and later collapse epochs.
Intriguingly, a steep transition of $f_{\rm star}$ for
$v_{\rm circ}\la 20{\rm km s^{-1}}$ coincidentally 
lies on a line of nearly constant $\sigma$, i.e., 
$\delta\rho/\rho \approx 2\sigma$.
This means that in more than 95\% of the halos
with $v_{\rm circ}\la 20{\rm km s^{-1}}$, the star formation
is strongly suppressed by the early reionization. 

In the present simulation, the star formation is assumed to
proceed in local free-fall time. This leads to the physically maximal
star formation rate. Hence, the obtained stellar fraction in a halo 
should be regarded as a maximal one. If the star formation 
proceeds in a longer timescale, 
baryon gas in a halo with $v_{\rm circ}\la 20{\rm km s^{-1}}$
could be almost completely photo-evaporated after the reionization.
On the other hand, fluctuations with $v_{\rm circ}\ga 20{\rm km s^{-1}}$
is like to be impervious to the photo-evaporation, 
since the gravitational potential is deep enough to retain the ionized gas. 

To examine the the dependence of results on assumed UVB intensity ($I_{21}$),
we also perform several runs with $I_{21}=1$. 
In this case, a steep transition of $f_{\rm star}$ is slightly shifted to
higher $\sigma$ as $\delta\rho/\rho \approx 2.5\sigma$.
Thus, it turns out that the results are not strongly dependent on the 
UVB intensity.
Such insensitiveness may be understood by the weak dependence
of the self-shielding on the UV intensity seen in equation 
(\ref{nshield}).

Finally, to check the numerical effects,
we analyze the convergence of the runs with changing
the mass resolution and softening length.
In Figure \ref{fig2}, the left panel shows 
the final stellar fraction ($f_{\rm star}$) in runs with 
$M_{\rm vir} = 6\times 10^7 M_\odot$ and $z_{\rm c}\simeq 10$, 
while the right panel is with 
$M_{\rm vir} = 6\times 10^6 M_\odot $ and
$z_{\rm c}\simeq 15$. 
They are at the transition region of $f_{\rm star}$ in Figure \ref{fig1}. 
The horizontal axis in Figure \ref{fig2} is the
softening length $\epsilon$ for gravity. 
Four different curves correspond to
the different numbers of particles used in the simulations. 
We can see that $f_{\rm star}$ almost converges, 
if $N_{\rm SPH}\ga 2^{15}$ and if $\epsilon \la 10-20{\rm pc}$. 
Thus, the present simulation with $\epsilon=20{\rm pc}$ and 
$N_{\rm SPH}=2^{15}(=N_{\rm DM})$ is unlikely to suffer 
from numerical effects. 

\section{Discussion}
\label{Discussion}

The present numerical simulations predict strong negative feedback 
on the formation of dwarf galaxies with 
$v_{\rm circ} \la 20 {\rm km~s^{-1}}$. 
Based on $f_{\rm star}$, we can make a rough estimation of the mass-to-light 
ratios ($M/L$) for finally formed galaxies.
If we assume $M_{\rm star}/L=3$ for stars in solar units,
$f_{\rm star}=0.01$ corresponds to $M_{\rm vir}/L =2.6 \times 10^{3}$
and $f_{\rm star}=0.1$ to $M_{\rm vir}/L= 2.6 \times 10^{2}$.
But, these values cannot be compared directly with the observed $M/L$
of satellite galaxies, because
a quite large fraction (typically more than 90 percent) of dark matter halos 
of satellites can be tidally stripped, as shown by \citet{Kravtsov04}.
Thus, the eventual $M/L$ of satellites is likely to decrease by a factor of 10,
e.g. $M/L =2.6 \times 10^{2}$ for $f_{\rm star}=0.01$.
Local Group dwarf galaxies (dSphs and dIrrs)
exhibit a wide range of $M/L$, which is from
a few up to $\approx 100$ \citep{vdB99,Mateo98,Hirashita98}.
Hence, the formed galaxies with $f_{\rm star} \ga 0.01$ may 
account for the Local Group dwarf galaxies. 
The present simulation predicts that only a few percent
of fluctuations result in $f_{\rm star} \ga 0.01$,
if $v_{\rm circ}\la 20{\rm km s^{-1}}$.
If the star formation rate is lower, the probability is
reduced further. Hence, such intrinsically low-mass halos
may be too few to account for all Galactic satellites.

One possibility to reconcile this discrepancy could be
a model suggested by \citet{Kravtsov04}.
They found that 10\% of small halos
originate in considerably larger systems with $\ga 10^9 M_\odot$,
which survived the tidal stripping, and 
suggest that the Galactic satellites are descendants of relatively 
massive systems which formed at higher redshifts.
In our simulation, systems larger than $10^8 M_\odot$
are not subject to the UVB feedback.
Thus, their model seems viable to account for the number of
Local Group dwarf galaxies.

\acknowledgments
We are grateful to the referee, who provided helpful comments on
this paper.
We thank A. Ferrara, T. Kitayama, K. Omukai, K. Wada, N. Yoshida,
and S. White for
stimulating discussion.  The HMCS has been developed in a project which
Center for Computational Physics, University of Tsukuba propelled in the course of JSPS
Research-for-the-Future program of Computational Science and
Engineering.
The analysis has been made with computational facilities 
at Center for Computational Sciences 
in University of Tsukuba and Rikkyo University. 
We acknowledge Research Grant from Japan Society for the Promotion of
Science (15740122:HS, 15340060:MU).




\setcounter{figure}{0}

\clearpage
\begin{figure}
\plottwo{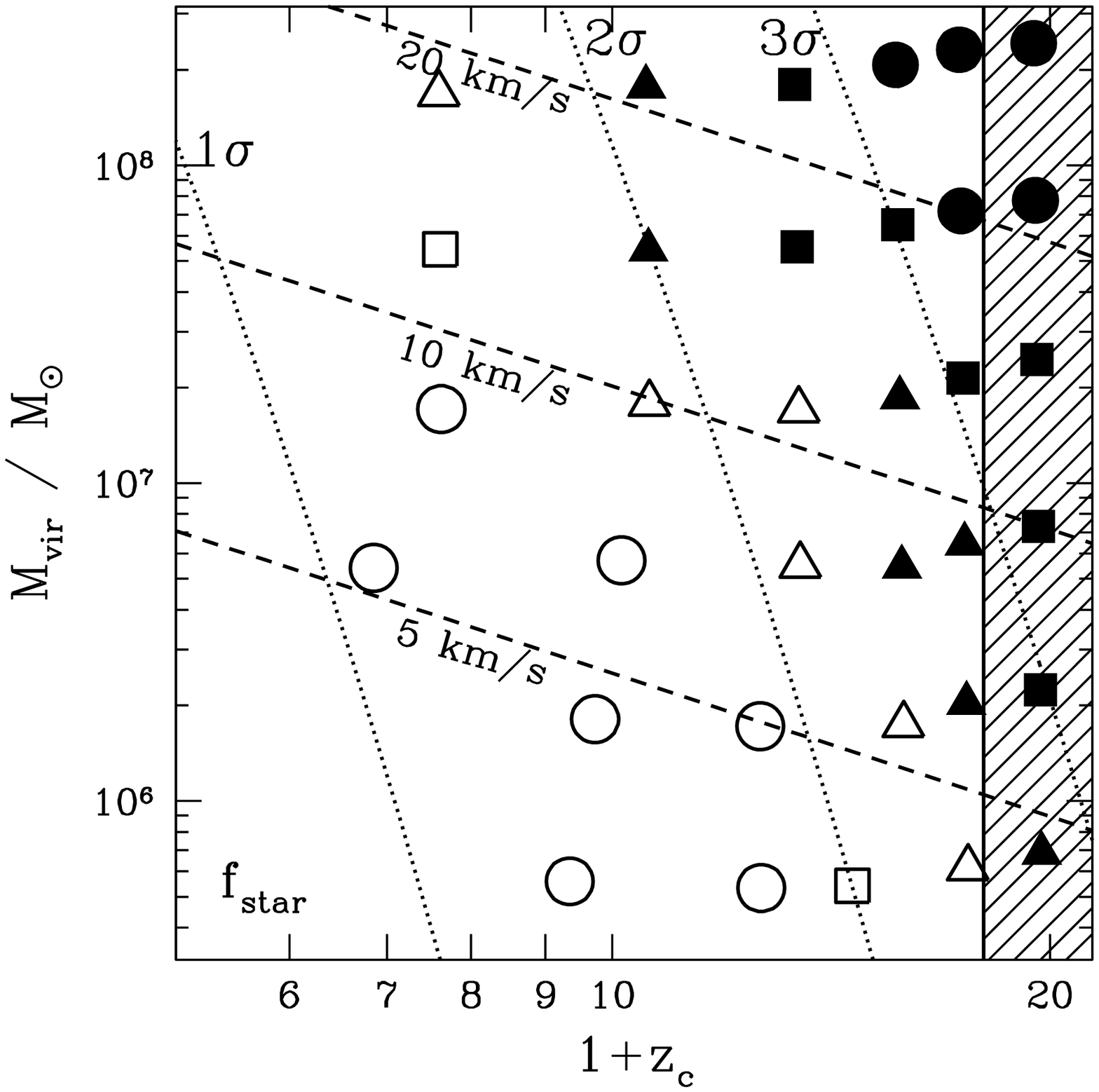}{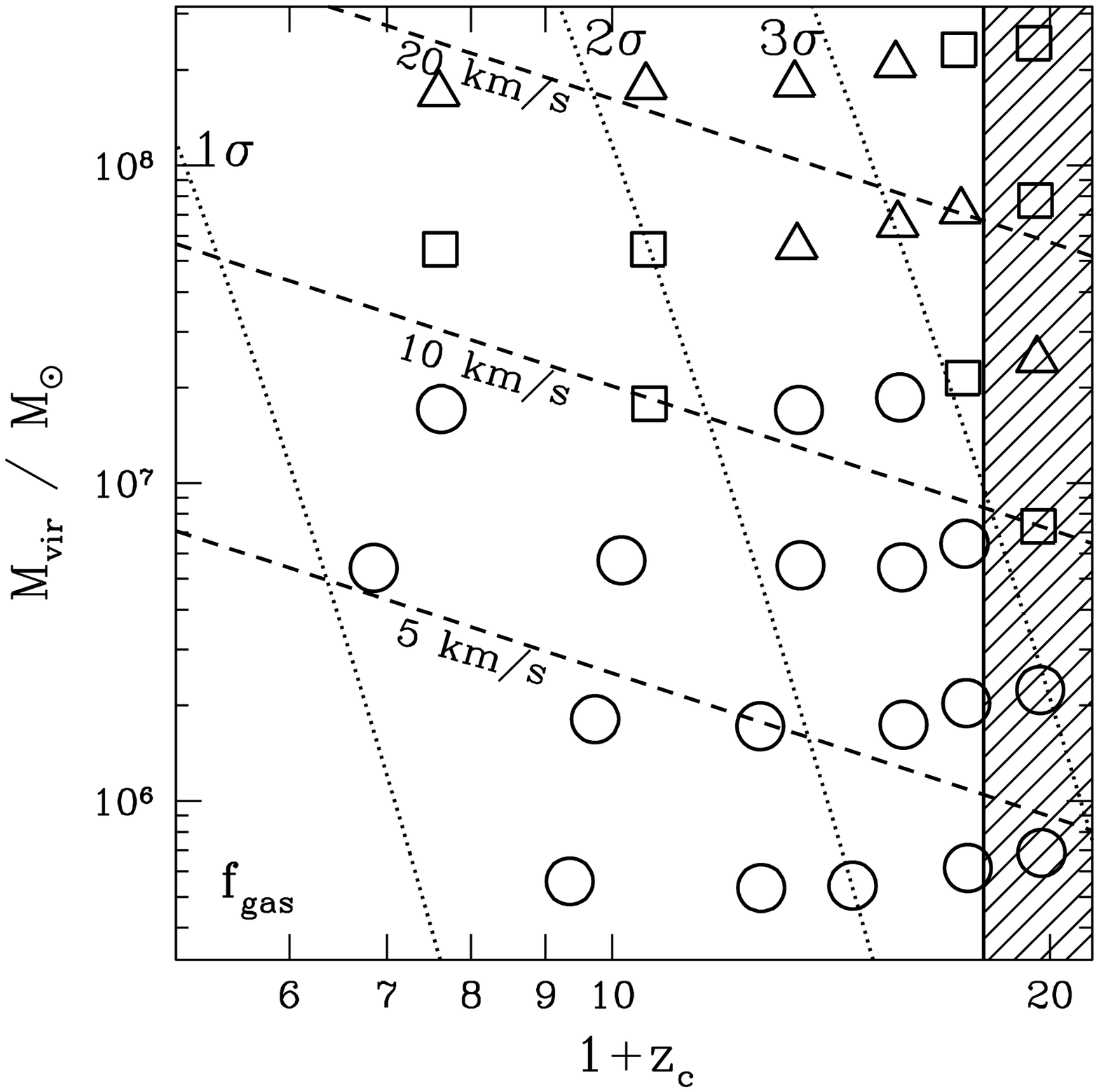}
\caption[dummy]{Summary of numerical runs are shown 
as a function of the collapse epoch $z_{\rm c}$ and
the virialized halo mass $M_{\rm vir}$.
The left panel shows the fraction of the final mass in stellar component
to the initial baryon mass, while the right panel is the final fraction of gas. 
In both panels, the fraction $f$ is depicted by different symbols;
{\it filled circles} are $f>0.9$, {\it filled squares} $0.5<f<0.9$,
{\it filled triangles} $0.1<f<0.5$, {\it open triangles} $0.01<f<0.1$,
{\it open squares} $10^{-3}<f<0.01$, and {\it open circles} $f<10^{-3}$.
Dotted lines
represent the collapse epoch of halos formed from $1\sigma, 2\sigma$ and
$3\sigma$ CDM density fluctuations. 
Dashed lines denote the constant circular velocities of 5 km/s, 10km/s, and 20km/s. 
}\label{fig1}
\end{figure}

\begin{figure}
\plottwo{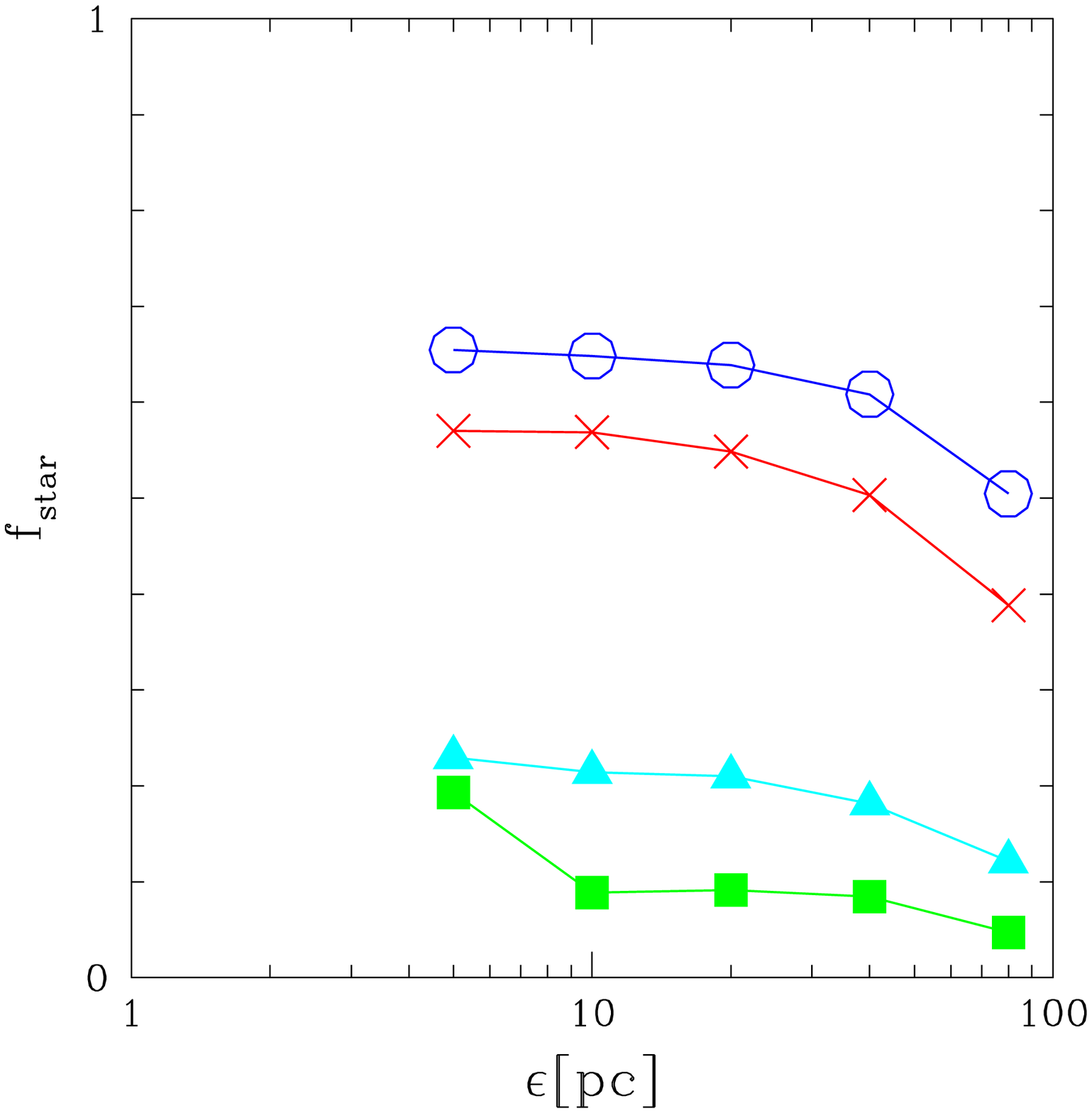}{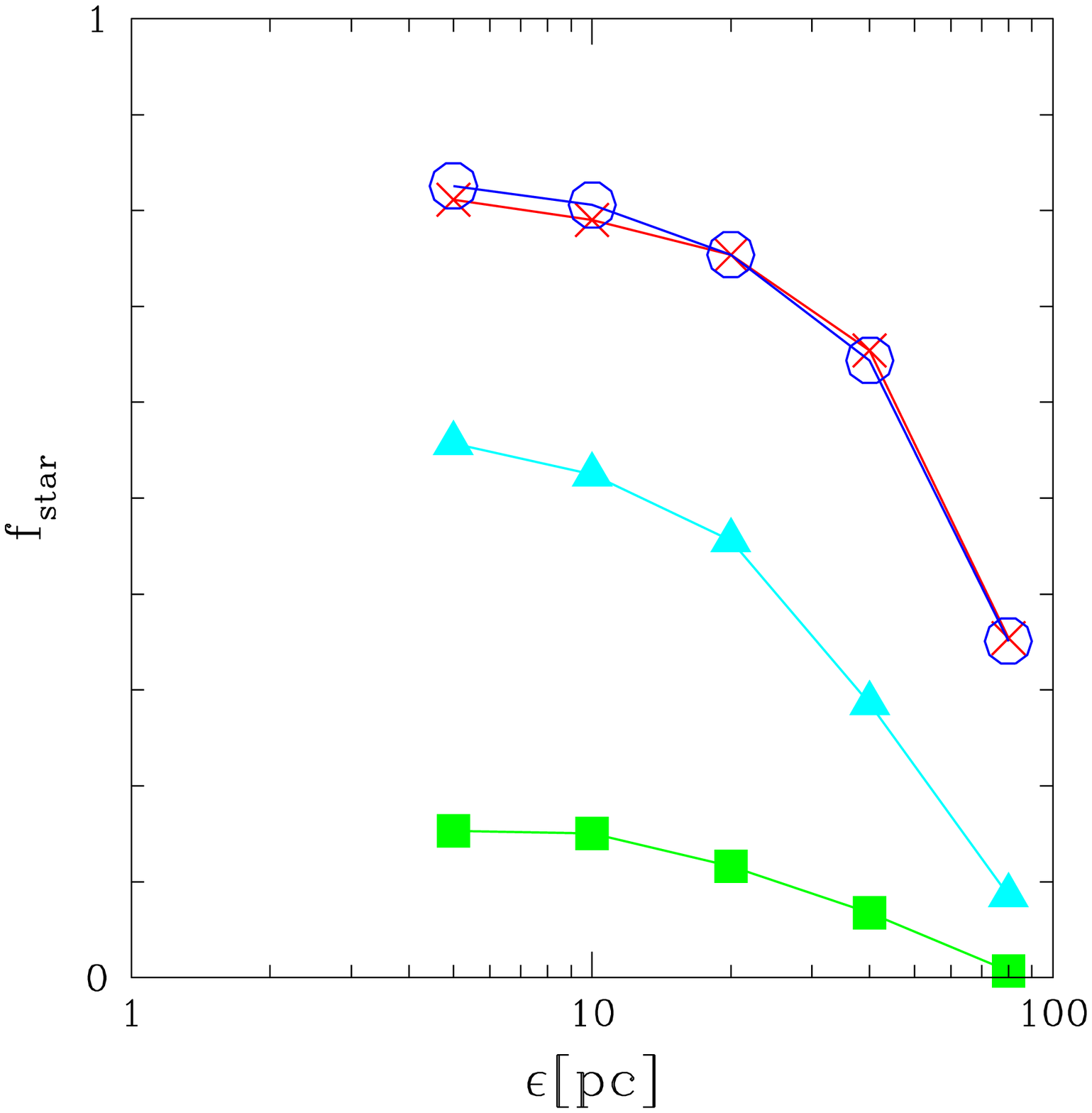}
\caption[dummy]{Convergence check of the numerical simulations.
 The left panel shows the convergence of runs with 
$M_{\rm vir}=6\times 10^7 M_\odot and
 z_{\rm c}\simeq 10$, and the right panel is for
$M_{\rm vir}=6\times 10^6 M_\odot and z_{\rm c}\simeq 15$. 
In both panels, vertical axes represent the final stellar mass fraction, 
and the horizontal axes show the softening length of the
 gravitational force. Four different curves represent the results with
 various number of particles: {\it circles} denote $N_{\rm SPH}=2^{17}$, 
{\it crosses} $N_{\rm SPH}=2^{15}$, {\it triangles} $N_{\rm SPH}=2^{14}$, 
and {\it squares} $N_{\rm SPH}=2^{13}$.
}\label{fig2}
\end{figure}

\end{document}